\documentclass[apj]{emulateapj}
\usepackage{epsfig}
\usepackage{color}

\shorttitle{Electron Shock Acceleration at Solar Flares}
\shortauthors{Guo and Giacalone}
\singlespace
\begin{document}

\title{Particle Acceleration at a Flare Termination Shock: \\ Effect of Large-scale Magnetic Turbulence}

\author{Fan Guo and Joe Giacalone}
\doublespace
\affil{Department of Planetary Sciences, University of Arizona, Tucson, AZ 85721, USA}

\email{guofan@lpl.arizona.edu}

\begin{abstract}
We investigate the acceleration of charged particles (both electrons and protons) at collisionless shocks predicted to exist in the vicinity of solar flares. The existence of standing termination shocks has been examined by flare models and numerical simulations \citep[e.g.,][]{Shibata1995,Forbes1986}. We study electron energization by numerically integrating the equations of motion of a large number of test-particle electrons in the time-dependent two-dimensional electric and magnetic fields generated from hybrid simulations (kinetic ions and fluid electron) using parameters typical of the solar flare plasma environment. The shock is produced by injecting plasma flow toward a rigid piston. Large-scale magnetic fluctuations -- known to exist in plasmas and known to have important effects on the nonthermal electron acceleration at shocks -- are also included in our simulations. For the parameters characteristic of the flaring region, our calculations suggest that the termination shock formed in the reconnection outflow region (above post-flare loops) could accelerate electrons to a kinetic energy of a few MeV within $100$ ion cyclotron periods, which is of the order of a millisecond. Given a sufficient turbulence amplitude level ($\delta B^2/B_0^2 \sim 0.3$), about 10\% of thermal test-particle electrons are accelerated to more than $15$ keV. We find that protons are also accelerated, but not to as high energy in the available time and the energy spectra are considerably steeper than that of the electrons for the parameters used in our simulations. Our results are qualitatively consistent with the observed hard X-ray emissions in solar flares.
\end{abstract}

\section{Introduction}

Solar flares are observed to be strong sources of energetic charged particles \citep{Aschwanden2002}. The release of magnetic energy by magnetic reconnection is thought to be the driving process \citep{Masuda1994}. While several mechanisms have been proposed to explain the acceleration of charged particles in flares \citep[see review by][and references therein]{Miller1997}, there is still no general consensus and this remains an unsolved problem. Recent hard X-ray observations of the nonthermal electron bremsstrahlung emission by \emph{Reuven Ramaty High Energy Solar Spectroscopic Imager} \citep[\emph{RHESSI; }][]{Lin2002} have provided more details of electron acceleration in solar flares. The observations indicate a large fraction of released energy resides in high-energy electrons accelerated soon after the flare. Hard X-ray sources above the top of magnetic loops have been detected \citep[e.g.,][]{Masuda1994,Krucker2010}, providing important clues to the acceleration process. For example, the loop-top source recently reported by \citet{Krucker2010} shows a large number of electrons ($> 5\times 10^{35}$) are accelerated to more than 16 keV and the highest energy reaches $\sim$ MeV. Accelerated ions have also been observed in solar flares and tend to correlate with energetic electrons \citep{Shih2009}. Since the observed hard X-ray source requires a very efficient acceleration, explaining how such a large number of electrons (probably also ions) are accelerated to high energy poses a challenge to theoretical astrophysics.

\begin{figure}
 \begin{center}
\includegraphics[width=40pc]{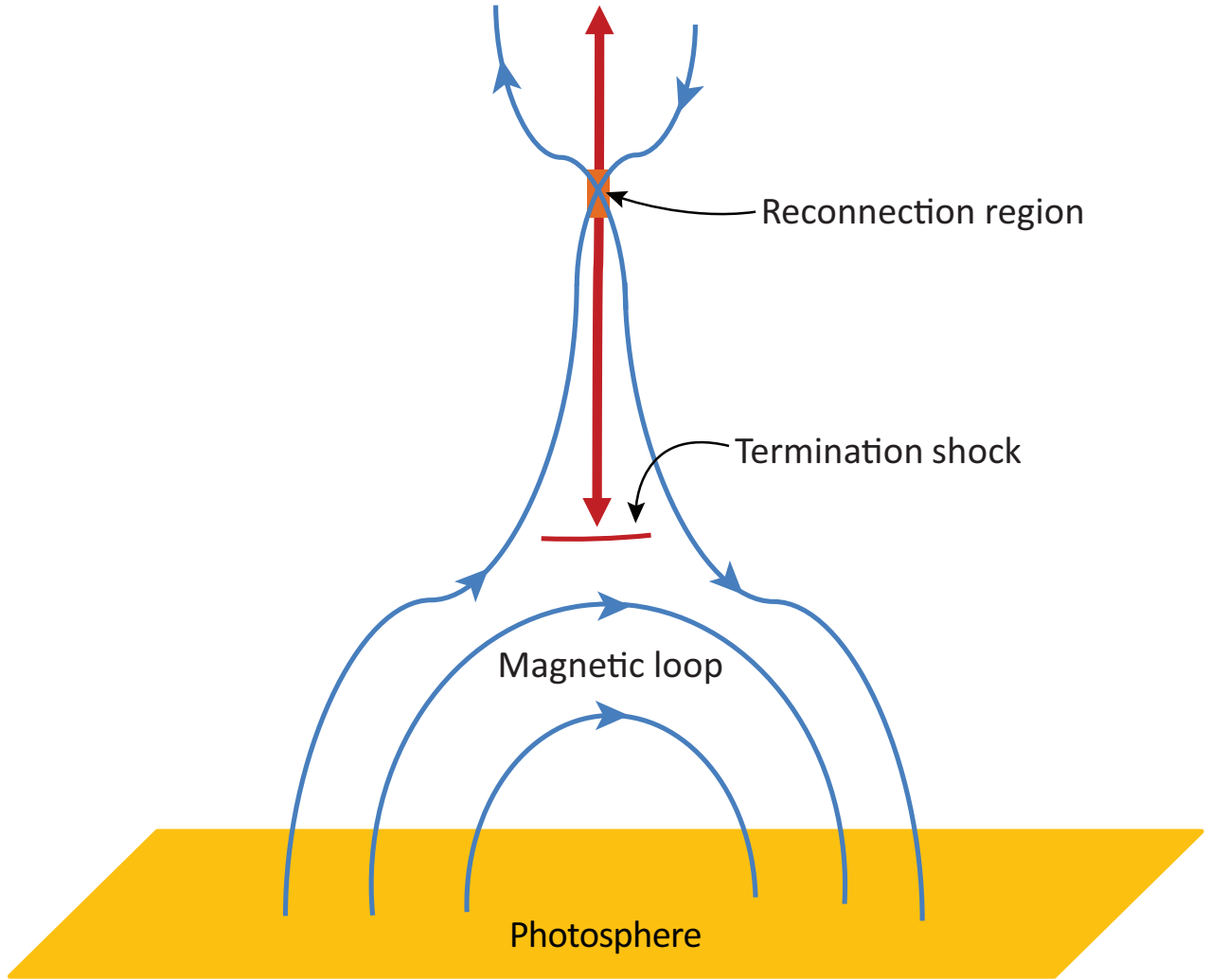}
 \caption{A cartoon illustration of geometry of flare termination shock.}
 \label{cartoon}
 \end{center}
 \end{figure}

The existence of fast shocks in the reconnection outflow region has been predicted in flare models \citep[e.g.,][]{Shibata1995} and numerical simulations \citep[e.g.,][]{Forbes1986,Forbes1988,Shiota2003,Workman2011}. Using MHD numerical simulations, \citet{Forbes1986} studied the formation of a standing termination shock when a high-speed jet driven by reconnection encounters a closed magnetic loop. The geometry of the flare termination shock can be represented by Figure \ref{cartoon}. The high-speed jet created in the reconnection outflow region collides with the top of the magnetic loop and produces a fast-mode, standing termination shock. The resulting flare termination shock has a unit normal to its surface that points nearly perpendicular to the magnetic field. This is a perpendicular shock (i.e., the angle between the upstream magnetic field and shock normal vector $\theta_{Bn} = 90^\circ$). \citet{Forbes1986} predicts the existence of this shock with a compression ratio of 2.0 and an upstream Mach number as high as 2.3. A recent study by \citet{Workman2011} shows similar results. The observational evidence of the existence of flare shocks has been presented by \citet{Aurass2002}.

Fast-mode, collisionless shocks are known to be efficient accelerators of charged particles \citep{Blandford1987}. The theory of diffusive shock acceleration \citep[e.g.,][]{Axford1977,Bell1978} predicts shock-accelerated particles have a power-law distribution, which is often seen in solar energetic particle events.
Energetic electrons are frequently observed to be associated with collisionless shocks with $\theta_{Bn} > 45^\circ$ \citep[e.g.,][]{Fan1964,Anderson1979,Tsurutani1985,Simnett2005,Oka2006}.
Electron acceleration by oblique shocks has been considered by a number of authors \citep{Holman1983,Wu1984,Leroy1984,Krauss-Varban1989a,Krauss-Varban1989b,Guo2010,Riquelme2011,Matsukiyo2011}. In the scatter-free limit, the acceleration of electrons at a planar quasi-perpendicular shock is often referred to as fast-Fermi acceleration or shock drift acceleration. It can be demonstrated that these two mechanisms are the same process in two different frames of reference \citep{Krauss-Varban1989b}. Shock drift acceleration at a flare termination shock in the reconnection outflow region has been considered by a number of authors \citep[e.g.,][]{Mann2009,Warmuth2009}. However it is commonly known that in this process both the fraction and attainable energy of the accelerated particles are limited \citep[e.g.,][]{Ball2001}. The acceleration and scattering of low-rigidity particles, such as electrons is thought to be a problem. This is commonly referred to as ``injection problem". For low-rigidity ions, the injection problem is better understood. It is usually thought that the ion injection at quasi-parallel shocks is easier than that at quasi-perpendicular shock (see discussions by \citet{Lembege2004} and more recent numerical simulation studies by \citet{Yang2009} and \citet{Wu2009}). However, numerical simulations that consider large-scale pre-existing  magnetic turbulence suggests there is no injection problem \citep{Giacalone2005a,Giacalone2005b}.
There has been some works on facilitating particle acceleration at flare termination shocks. \citet{Tsuneta1998} considered electron heating by a slow-shock pair as a pre-energization process. \citet{Somov1997} considered the role of plasma heating and collapsing magnetic trap at reconnecting magnetic field lines.

One of the recent advances in our understanding of electron acceleration at shocks is the consideration of nonuniform effects such as shock ripples and magnetic turbulence. The simulations by \citet{Burgess2006} and \citet{Umeda2009} showed that small-scale shock ripples can be important in scattering the electrons and facilitating the acceleration. \citet{Savoini2010} studied the motions of electrons in the vicinity of a shock using two-dimensional full particle simulations. They found the effects of shock reformation and shock rippling are important for non-adiabatic electron heating at collisionless shocks. Recently, \citet{Jokipii2007} proposed a mechanism to solve the injection problem which does not require strong pitch-angle scattering from small-scale fluctuations. The fast-moving electrons can move along meandering magnetic field lines and travel back and forth across the shock front, and, therefore gain energy from the difference between the upstream and downstream flow velocities. Using self-consistent hybrid simulation combined with test-particle simulation for electrons, \citet{Guo2010} have found efficient electron acceleration at perpendicular shocks moving through a plasma containing large-scale pre-existing upstream magnetic turbulence. The turbulent magnetic field leads to field-line meandering which allows the electrons to get accelerated at the shock front multiple times, like in the Jokipii \& Giacalone picture. The shock surface becomes irregular on a variety of spatial scales from small-scale ripples due to ion-scale plasma instabilities \citep{Lowe2003}, to large-scale structures caused by the interaction between the shock and upstream large-scale turbulence \citep{Giacalone2008,Lu2009}. The rippled surface of the shock front also contributes to the acceleration by mirroring electrons between the ripples. The observational evidence of these large-scale ripples have been shown by a number of authors \citep{Neugebauer2005,Bale1999,Pulupa2008,Koval2010}. These results, along with the previous work on acceleration of ions \citep{Giacalone2005a,Giacalone2005b}, suggest that large-scale turbulence has an important effect on the acceleration of both electron and ions at shocks, which is consistent with the correlation between ions and electrons in solar energetic particle events \citep[e.g.,][]{Cliver2009,Haggerty2009}. This mechanism especially favors perpendicular shocks, particularly for electrons, and could explain the observed high-energy particles in solar flares as well.

In this paper we present results from a combination of hybrid simulations and test-particle simulations to study the electron and proton energization at a shock with parameters similar to the flare termination shock inferred by previous works in the existence of upstream magnetic fluctuations. We consider the nonlinear modification of a planar shock front by upstream Alfvenic fluctuations and its effect on electron acceleration. Although the plasma waves and turbulence in the reconnection outflow region could be considerably different from our simplified model, some intrinsic characteristics of this interaction, such as the braiding of magnetic field lines and shock rippling are still preserved. Our study suggests the acceleration of electrons at a flare termination shock is rapid and efficient. A large fraction of the initial thermal electrons is accelerated to hundreds of keV and even reaches MeV energies in a very short time. This indicates collisionless shocks may play an important role in particle acceleration in solar flares. In section 2 we describe the numerical method used in this paper. Section 3 discusses the simulation results. In section 4 we discuss the implication of our results and draw conclusions.

\section{Numerical Method}

 We use a combination of two-dimensional hybrid simulations and test-particle electron simulations to study the energization of electrons at the flare termination shock. In the hybrid simulation model \citep{Winske1985}, the ions are treated kinetically and the background thermal electrons are usually treated as a massless, charge neutralizing fluid. The electrons are solved by using electron fluid momentum equation (generalized Ohm's law).  This approach is well suited to resolve the kinetic physics of ions and has been successfully applied to study the physics of collisionless shocks. We consider a two-dimensional Cartesian geometry in which all physical quantities are functions of two spatial variables $x$ and $z$. All vectors have components in three directions $(\hat{x}, \hat{y}, \hat{z})$. The size of the simulation domain $L_x \times L_z$ for each case is listed in Table \ref{table}.  The flare termination shock is modeled by injecting plasma continuously from one end $(x=0)$ of the simulation box and colliding with a rigid wall at the other end $(x=L_x)$. A simplified one-dimensional fluctuating magnetic field $\textbf{B}(z,t) = \delta \textbf{B}(z,t) + B_{10}\hat{z}$ is present in the simulation box at $t=0$ and also injected continuously at the $x=0$ boundary with the plasma flow during the simulation, where $B_{10}$ is the averaged upstream magnetic field. The fluctuating component $\delta \textbf{B}(z,t)$ consists of an equal-partition of right- and left-hand circular polarized, forward and backward parallel-propagating plane Alfven waves. The amplitude of the fluctuations at the wave number $k$ is determined from a Kolmogorov power spectrum:

$$
P(k) \propto \frac{1}{1 + (k L_c)^{5/3}}
$$

 \noindent where $L_c$ is the coherence length. The total variance, which is the integral of $P$ over all $k$, in each case is listed in Table 1. In all the cases, we take the turbulence coherence length to be equal to the box size in the $z$ direction $L_c = L_z$. We expect large-scale magnetic turbulence to exist in the reconnection outflow plasma. These fluctuations can be triggered by reconnection, foot-point motion or other processes (see \citeauthor{Krucker2008}, \citeyear{Krucker2008}, for a detailed discussion).  Although the simplified form of magnetic-field fluctuations we use may not be representative of the reconnection outflow plasma, any turbulence with large variances should allow strong field-line wandering, which is essential in our particle acceleration model.

In order to produce a low Mach number shock, as predicted by other numerical simulations, the inflow Alfven Mach number is taken to be $M_{A0} = 1.0$. After reflection at the right boundary, this produces a shock with an average Mach number of about $2.0$ in the shock frame, consistent with the flare termination shock predicted by previous MHD simulations \citep{Forbes1986,Workman2011}. The grid sizes are $\Delta x = \Delta z = 0.5
c/\omega_{pi}$ and the time step is taken to be $\Delta t = 0.01 \Omega_{ci}^{-1}$, where $c/\omega_{pi}$ is the ion inertial length and $\Omega_{ci}^{-1}$ is the ion cyclotron period. The ion and electron plasma betas, $\beta_i$ and $\beta_e$, respectively, are both taken to be $0.03$ and the ratio between light speed and upstream Alfven speed $c/V_{A0} = 410$, which roughly corresponds to an initial situation with proton and electron temperatures $T_i = T_e = 2\times 10^6 K$, number density $n = 8\times 10^9 cm^{-3}$ and magnetic field $B_0 = 30G$, similar to constraints from observations \citep{Krucker2010}. For these parameters, the average upstream speed in shock frame of reference (also the jet outflow speed) is about $1460 km/s$. This speed is measurable because the outflow plasma from reconnection is moving at the Alfven speed. The estimate from observations \citep{Tsuneta1997} is roughly consistent with this value.

Since in the hybrid model the electrons are treated as a massless fluid, the test-particle simulations are needed to study their kinetic behavior. In the second step, we integrate the relativistic equations of motion for an ensemble of test-particle electrons in the two-dimensional time-dependent electric and magnetic fields obtained in the hybrid simulations. These test-particle electrons are treated as a different part from the electron fluid in the hybrid simulations. We use a second-order spatial interpolation and linear temporal interpolation to get the field at the particle position. Initially we release 1.6 million test electrons which have a Maxwellian distribution with $T_e = 2.0 \times 10^6 K$ in the upstream frame. The test electrons are released upstream of the shock at the time $\Omega_{ci} t = 30$ when the shock is fully formed and far from the boundaries. They are released uniformly in space over all values of z and between $x = 350 c/\omega_{pi}$ and $x = 450 c/\omega_{pi}$. We implement the so-called Bulirsh-Stor method to integrate the trajectories of the electrons \citep{Press1986}. The method is highly accurate and fast when the fields are smooth on the scale of particle gyroradius. The algorithm uses an adaptive method based on the evaluation of the local truncation error, which is essential to reduce numerical error when electrons experience rapid field variations. The time step is allowed to vary between $10^{-3}$ and $0.2\Omega_{ce}^{-1}$ and the ratio $\Omega_{ce}/\Omega_{ci}$ is taken to be the realistic value $1836$. The boundary condition in the z-direction is taken to be periodic. The simulation domain in the x-direction is large enough so that no test-particle electrons escape from the system. The readers are referred to earlier works using this method for additional numerical details \citep{Krauss-Varban1989a,Burgess2006}. Strictly speaking, this test-particle simulation is only valid when the influence of the accelerated electron to the background fluid is negligible. However, in the end of the simulation, the electron distribution considerably departs from the initial Maxwellian due to the energization process at the shock front. The effect of the non-thermal electrons on the long-term evolution of the termination shock, which is not included in our calculation, may become significant. We also note that in our two-dimensional simulation (or any simulation with at least one ignorable coordinate), the motion of charged particle is tied to their original field lines \citep{Jokipii1993}. The effect of particles moving off their field lines of force could be important to this process, and requires fully three-dimensional simulations, which is beyond our available computing resources.

\section{Simulation Results}

The shocks in our study have low Mach numbers and high upstream inflow speed (corresponding to high speed jet from the reconnection outflow region). The plasma in the solar corona is hot and has a low beta (strong magnetic field), which is considerably different from that in interplanetary space. We focus on the modification of the shock surface by upstream Alfvenic fluctuations and its effect on the acceleration of electrons. Table \ref{table} lists some key parameters for all the simulation runs including the size of the simulation domain, the variance of the injected magnetic fluctuation, and the fraction of electrons whose energy is more than $15 keV$ at the end of simulation (also see Figure 5). For runs 1--4 we consider the effect of different variances of magnetic turbulence. The turbulence variances range from 0.0 to 0.3 and the size of the simulation domain is $L_x \times L_z = 500 c/\omega_{pi}\times 400 c/\omega_{pi} (1.27 km\times 1.02 km) $ for these four cases. For runs 5--7, the magnetic variances are the same as runs 2--4, but the size of the simulation box is changed to $L_x \times L_z = 500 c/\omega_{pi}\times 800 c/\omega_{pi} (1.27 km\times 2.03 km)$ to examine the effect of changing the coherence length of the turbulence which, in our case, is governed by the size of the box. In flare environment, strong large-scale Alfvenic magnetic fluctuations can be triggered by reconnection, and cascade to small scales. This process is usually assumed to be the source of magnetic turbulence required in many acceleration models \citep[e.g.,][]{Miller1996,Petrosian2004}. Since the size of our largest simulation domain is still much smaller than the observed hard X-ray emission region ($L \sim 10^3 km$), we do not consider a realistic geometry of flare termination shock but approximate it locally as a perpendicular shock that propagates into a plasma containing magnetic fluctuations.

\begin{figure}
\centering
\begin{tabular}{cc}
\epsfig{file=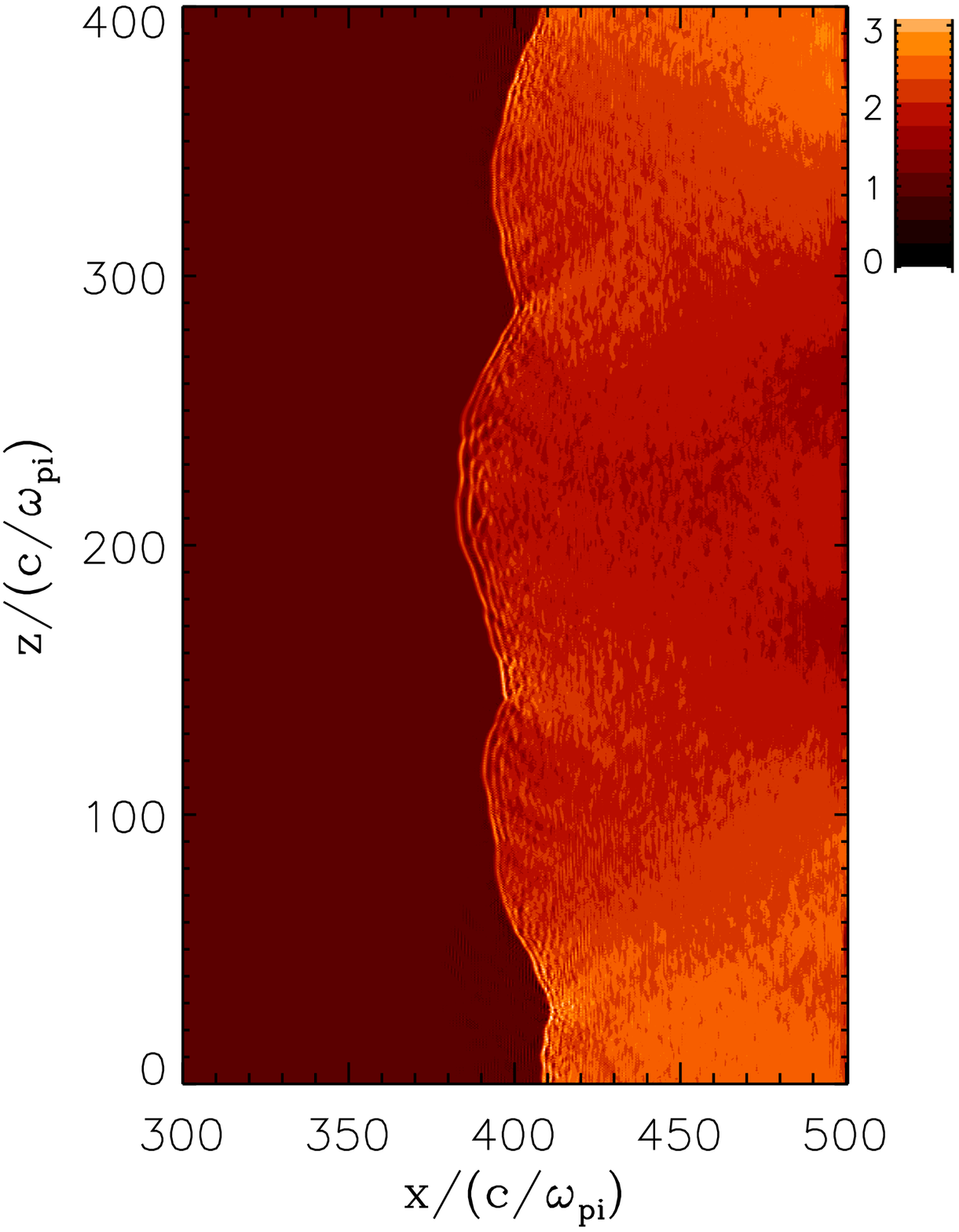,width=16pc,clip=} &
\epsfig{file=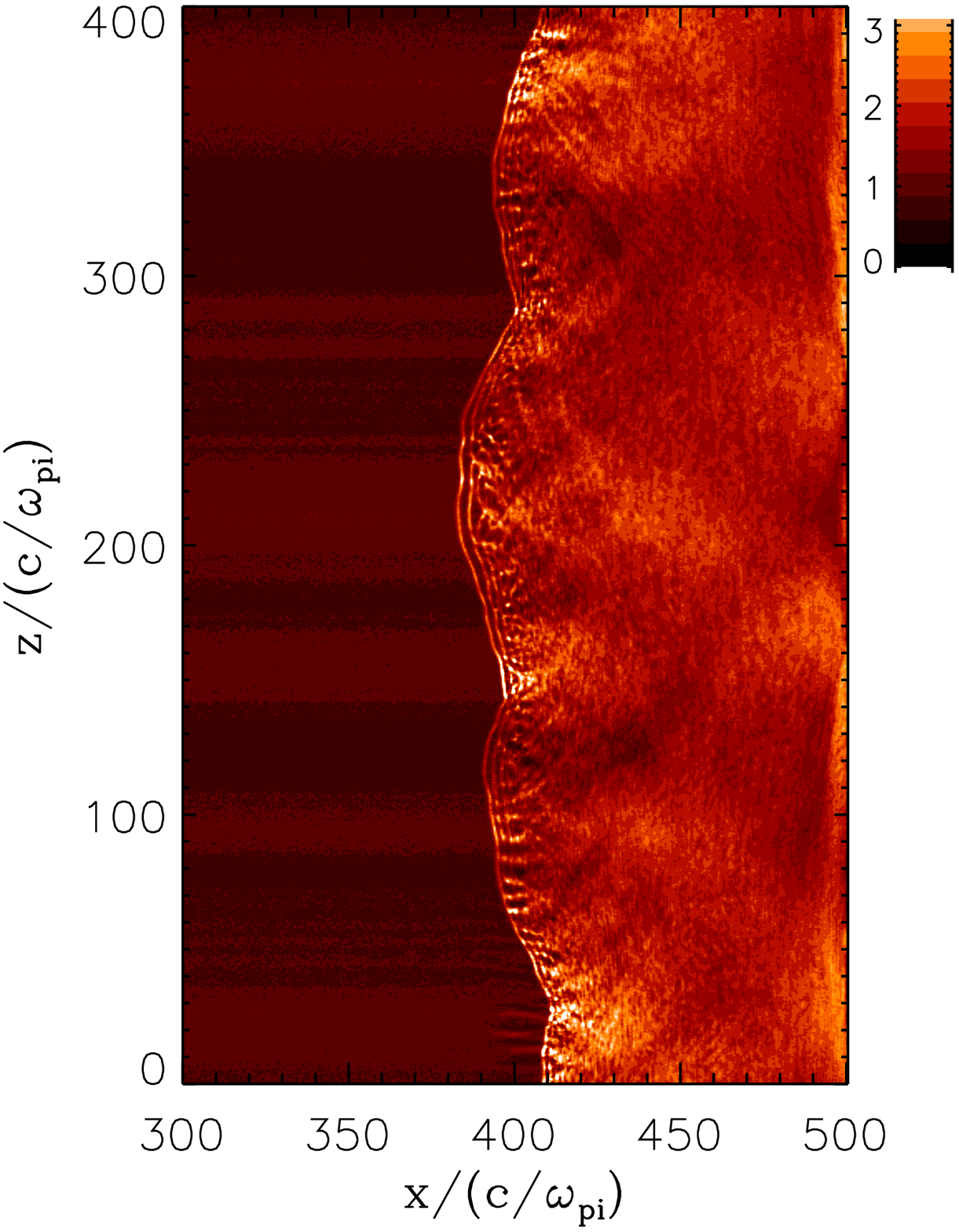,width=16pc,clip=}
\end{tabular}
\caption{The color-coded images of (a) magnetic field $B_z$, (b) ion density $n_p$ for run 3 at $\Omega_{ci} t=110.0$}
\label{field}
\end{figure}

Figure \ref{field} shows (a) the $z$-component of the magnetic field and (b) the ion number density $n_p$, in a color-coded representation from run 3 at $\Omega_{ci} t = 110.0$. The magnetic field and plasma density have been normalized using the average upstream magnetic field $B_{10}$ and in-flow density $n_0$. The averaged Alfven Mach number in the shock frame is about 2.0 and the average compression ratio is about 2.1. As noted in the earlier works of \citet{Giacalone2005b} and \citet{Guo2010}, the shock surface becomes distorted due to the interaction between the shock front and the upstream turbulence. Meandering magnetic field lines cross the shock front at various locations along the shock, which allows the electrons to cross and/or get reflected at the shock front multiple times. The shock-front rippling has also been shown to contribute to particle acceleration by mirroring electrons between ripples \citep{Guo2010}.

\begin{figure}
 \begin{center}
\includegraphics[width=20pc]{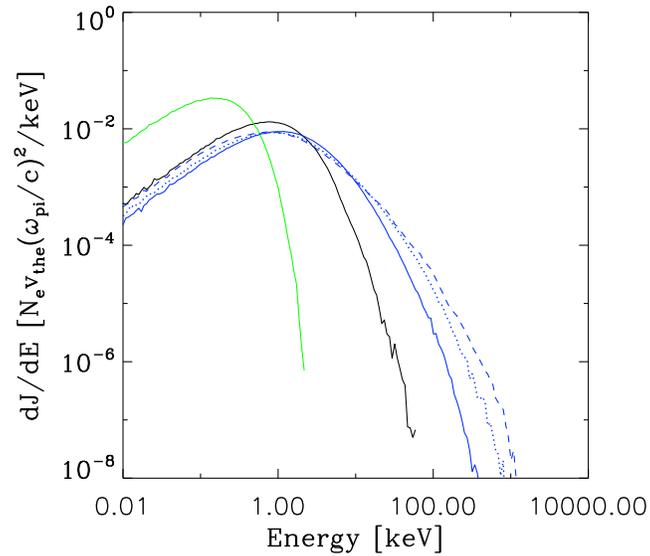}
 \caption{The energy spectra of electrons at the end of the simulation ($\Omega_i t = 130.0$). The energy spectra are normalized using $N_e v_{the} (\omega_{pi}/c)^2/keV$, where $N_e$ is the total number of electrons used in the simulations and $v_{the}$ is the initial electron thermal speed. The green solid line shows the initial distribution of thermal electrons in the upstream region. The black solid line displays the energy distribution for all the electrons in downstream region at the end of simulation for run 1.  The blue solid, dot and dashed lines represent results from runs 2, 3, and 4, respectively.}
 \label{spectrum1}
 \end{center}
 \end{figure}

 Figure \ref{spectrum1} presents the energy spectra,  $dJ/dE \ vs.\ E$, of electrons. The energy spectra are normalized using $N_e v_{the} (\omega_{pi}/c)^2/keV$, where $N_e$ is the total number of electrons used in the simulations and $v_{the}$ is the initial electron thermal speed.  The green solid line shows the initial distribution of thermal electrons in the upstream region. The black solid line displays the energy distribution for all the electrons downstream of the shock at the end of simulation ($\Omega_{ci} t = 130.0$) for run 1. In the case of no pre-existing fluctuation the electron energization is primarily due to heating and nonthermal acceleration within the shock layer. The resulting electron spectrum is broadened compared to the distribution incident on the shock and has a power-law-like tail extending to high energy. This is similar to previous work and confirms the effect of ion-scale ripples on the electron energization at the shock transition \citep{Burgess2006}. The effective electron kinetic temperature jump across the shock including the superthermal distribution is about $6$ times that of the upstream temperature. The electron temperature jump is about $40\%$ of the proton temperature jump across the shock layer in our hybrid simulation. This is consistent with the theoretical prediction that the heating of electrons in fast shocks is less than that of ions \citep{Goodrich1984,Scudder1995} and the observational constraints from measurements at planetary bow shocks and interplanetary shocks \citep{Thomsen1987,Schwartz1988}.

 In the following we focus on the nonthermal acceleration of electrons at shocks after considering the pre-existing magnetic fluctuations. The blue solid, dot and dashed lines in Figure \ref{spectrum1} show energy distribution for all the electrons in the downstream region at the end of the simulation ($\Omega_{ci} t = 130.0$) for runs 2--4, respectively. At this time most of electrons are downstream of the shock so the energy spectra do not evolve any longer. It can be seen that the electrons are accelerated to high energy when the upstream magnetic turbulence is included. For higher variance of magnetic turbulence, there are more particles accelerated to high energy. For run 4 ($\delta B^2/B_0^2 = 0.3$), $9.8\%$ of electrons are accelerated to more than $15 keV$ by the end of the simulation. The efficient electron acceleration can be understood as stronger magnetic turbulence allows increased field-line meandering, and the electrons move along field lines of force that intersect the shock at several places which therefore allows for efficient energy gain at the shock front. Compared with earlier work \citep{Guo2010}, we find the maximum energy is proportional to the square of upstream inflow speed in shock frame. This is consistent with diffusive shock acceleration in which the energy gain is proportional to $V_{sh}^2$, where $V_{sh}$ is shock speed in upstream frame.

\begin{figure}
 \begin{center}
\includegraphics[width=20pc]{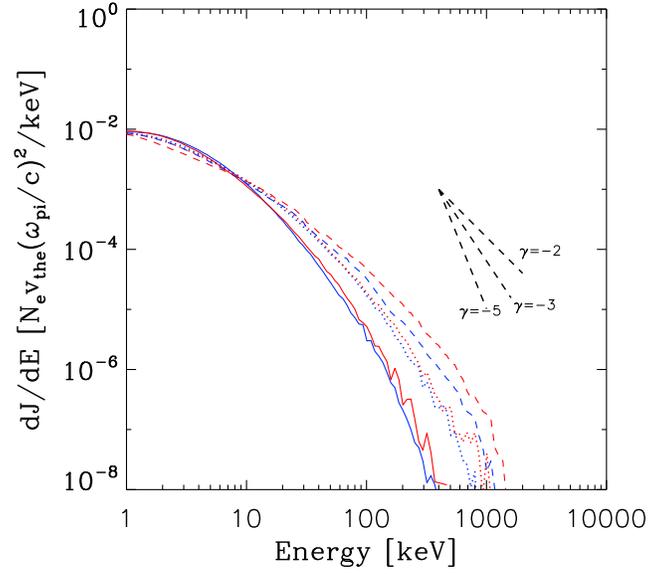}
 \caption{The energy spectra of electrons at the end of the simulation ($\Omega_i t = 130.0$). The energy spectra are normalized using $N_e v_{the} (\omega_{pi}/c)^2/keV$, where $N_e$ is the total number of electrons used in the simulations and $v_{the}$ is the initial electron thermal speed. The red solid, dot and dashed lines represent the energy distributions for all the electrons in downstream region at the end of simulation for runs 5, 6, and 7 respectively. The blue solid, dot and dashed lines represent results from runs 2, 3, and 4, respectively.}
 \label{spectrum2}
 \end{center}
 \end{figure}

In Figure \ref{spectrum2} we examine the effect of changing the coherence length of the magnetic turbulence and focus on the high energy part of the energy spectra. It shows results from runs 5--7 (red lines, $L_z = 800 c/\omega_{pi}$) along with corresponding runs 2--4 (blue lines, $L_z = 400 c/\omega_{pi}$). It is shown that for larger coherence length, the electrons reach higher energy and the spectral slope is flatter. The more efficient acceleration in runs 5--7 can be understood as the larger simulation domain in the direction of magnetic field allows more field line wandering normal to the shock ($\Delta X^2 \propto \Delta Z$, where $\Delta X$ is the field-line random walk normal to the averaged magnetic field and $\Delta Z$ is distance along the field) therefore the electrons move across the shock more easily. This dependence shows that long-wavelength fluctuations are important to accelerate electrons to high energy.

Figure 5 shows the relation between the amplitude of magnetic turbulence $\delta B^2/B_0^2$ injected in hybrid simulation and the percentage of electrons eventually accelerated to more than 15 keV. The triangles represent the case with $L_z = 400 c/\omega_{pi}$ whereas the squares represent the case with $L_z = 800 c/\omega_{pi}$. This result shows that for larger amplitude of magnetic turbulence, more electrons get accelerated to high energies. Once $\delta B^2/B_0^2$ reaches 0.3 or higher, about 10\% of electrons are eventually accelerated to more than 15keV. For the cases in which $L_z = 800 c/\omega_{pi}$, the simulations generally give slightly more electrons accelerated to high energies.

We also analyze the acceleration of protons, which are treated self-consistently in this problem (i.e., they are included in the hybrid simulation). Figure $6$ shows the downstream energy spectra of protons in shock frame at $\Omega_{ci} t = 130.0$ for runs 2--7, normalized using $N_p v_{thp} (\omega_{pi}/c)^2/keV$, where $N_p$ is the total number of protons used to plot the spectra and $v_{thp}$ is the initial proton thermal speed. Similar to Figure \ref{spectrum2}, the results from runs 5--7 are represented by red lines and the results from runs 2--4 are displayed using blue lines. The accelerated protons show a similar dependence on turbulence variance and coherence length to that of electrons. This dependence has been found previously in the case of higher Mach number shock and larger coherence length \citep{Giacalone2005b}. These results show both electrons and protons are efficiently accelerated. However, for the parameters we use, the slopes of the energy spectra of protons are considerably steeper than that of the spectra of electrons. This is probably due to the fact that fast moving electrons can interact with shock front more times than protons in the given time, which allows a more efficient acceleration.
Protons may need more time and larger spatial scales to reach an efficient acceleration \citep{Giacalone2005a}. Also, as shown in \citep{Giacalone2005b}, the effect of pre-existing fluctuations tend to be prominent for ions in the case of large coherence length ($\sim 4000 c/\omega_{pi}$). Exploring the relative acceleration between electrons and ions for different parameters will be a subject of future work.

\section{Discussion and Conclusions}

Understanding particle acceleration in solar flares is a challenge since only remote observations are available and it is hard to identify the main mechanism. While it is commonly thought that magnetic reconnection drives the energy release, the detailed physical process involved in accelerating the electrons and ions is still not clear. In this paper we studied electron acceleration at flare termination shocks whose existence in the flaring region has been predicted by numerical simulations and flare models. We find that electrons are rapidly and efficiently accelerated at such shocks in the presence of pre-existing magnetic fluctuations with parameters similar to that inferred by previous work \citep{Forbes1986,Workman2011}. The electrons are accelerated to a few $MeV$ in 100 ion cyclotron periods (of the order of a millisecond) and more than $10\%$ of thermal electrons are accelerated to over $15 keV$ given a sufficiently strong magnetic turbulence $(\delta B^2/B_0^2 \sim 0.3)$. We also show electron acceleration is more efficient for larger turbulence variance $\delta B^2/B_0^2$ and/or a larger turbulence coherence length $L_c$. Both of these indicate that large-scale field-line meandering plays an essential role in accelerating electrons at shock fronts. Our simulations suggest that when magnetic turbulence is present the flare termination shock could accelerate electrons to much higher energies than simple shock-drift acceleration \citep[e.g.,][]{Mann2009}. We note that a similar mechanism has been shown to efficiently accelerate ions and has similar dependence on the turbulence properties. This correlation between ions and electrons is actually commonly observed in solar energetic particle events \citep{Cliver2009}. For the parameters we use in our simulations, the accelerated protons have energy spectra steeper than that of electrons. This is different from the previous results for parameters similar to interplanetary space \citep{Giacalone2005b,Guo2010}. We note that these results are carried out for energies lower than the injection energy for diffusive shock acceleration and can be variable and depend on parameters such as the property of magnetic turbulence.

We also note for the situation we study, the resulting distribution of electrons is non-Maxwellian. The structure of collisionless shock may be considerably modified by the accelerated particles. While this effect is not considered in our test particle simulations, it may be important. A full particle simulation with similar parameters will also be useful to study the energy partition between electron and protons during the acceleration at the flare termination shock.
While our results are qualitatively consistent with the observed hard X-ray emissions in solar flares, other plasma effects in the flare region may need to be considered to directly compare with the observations in the future.

\section*{Acknowledgement}
This work was supported by NASA under grants NNX10AF24G and NNX11AO64G.

\singlespace
\bibliographystyle{apj}



\clearpage

\begin{table}
\begin{tabular*}
{0.65\textwidth}{cccc}
\hline
Run& $L_x (c/\omega_{pi}) \times L_z (c/\omega_{pi})$ & $\delta B^2/B_0^2$ & $\Gamma\% (E>15keV)$  \\
\hline
1  & $500\times 400$ & 0.0  & 1.3\\
2  & $500\times 400$ & 0.03 & 4.5\\
3  & $500\times 400$ & 0.1  & 8.0\\
4  & $500\times 400$ & 0.3  & 9.8\\
5  & $500\times 800$ & 0.03 & 4.9\\
6  & $500\times 800$ & 0.1  & 8.9\\
7  & $500\times 800$ & 0.3  & 11.9\\
 \hline
\end{tabular*}
 \caption{Some parameters for different simulation runs. The size of the simulation domain, the variance of injected magnetic fluctuation, and the fraction of electrons whose energy is more than $15 keV$ at the end of simulation.}
 \label{table}
\end{table}

  \begin{figure}
 \begin{center}
\includegraphics[width=20pc]{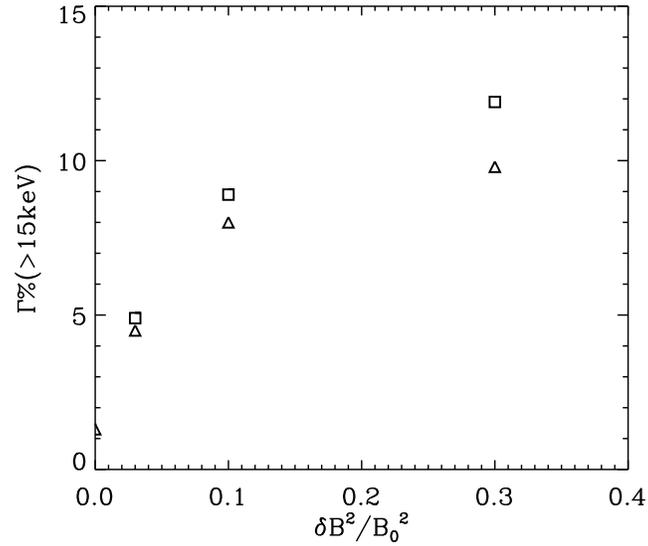}
 \caption{The relation between the turbulence amplitude $\delta B^2/B_0^2$ injected in hybrid simulation and the percentage of electrons eventually accelerated to more than 15 keV}
 \end{center}
 \end{figure}

 \begin{figure}
 \begin{center}
\includegraphics[width=20pc]{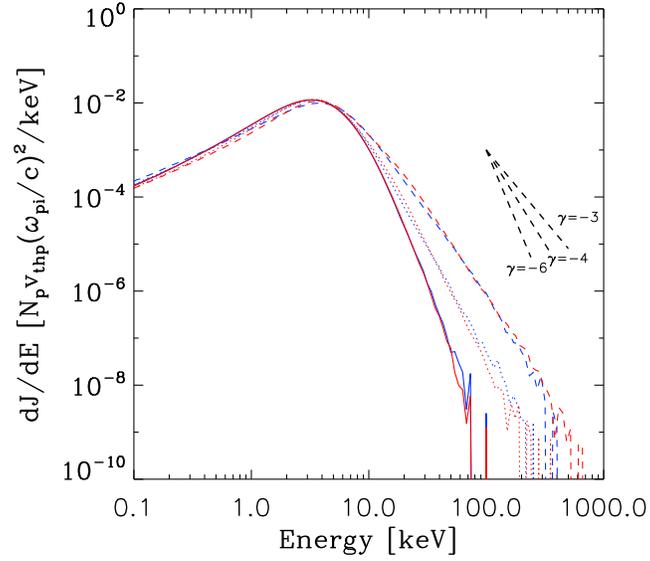}
 \caption{The energy spectra of protons downstream of the shock at the end of the simulation ($\Omega_i t = 130.0$), normalized using $N_p v_{thp} (\omega_{pi}/c)^2/keV$, where $N_p$ is the total number of protons used to plot the spectra and $v_{thp}$ is the initial proton thermal speed.  The red solid, dot and dashed lines represent the energy spectra for protons in downstream region at the end of simulation for runs 5, 6, and 7 respectively. The blue solid, dot and dashed lines represent results from runs 2, 3, and 4, respectively.}
 \label{figure5}
 \end{center}
 \end{figure}

\end{document}